\documentclass[amsmath,amssymb,pra,aps,showpacs,twocolumn,10pt,superscriptaddress,tightenlines]{revtex4-2}
\usepackage[colorlinks=true,citecolor=blue,linkcolor=red,urlcolor=blue]{hyperref}
\usepackage{graphicx}% Include figure files
\usepackage{dcolumn}% Align table columns on the decimal point
\usepackage{bm}% bold math
\usepackage[utf8]{inputenc}
\usepackage[T1]{fontenc}
\usepackage{lmodern}
\usepackage{cleveref}
\usepackage{color}
\usepackage{amsmath}
\usepackage{amssymb}
\usepackage{amsfonts}
\usepackage{latexsym}
\usepackage{multirow}
\usepackage{booktabs}
\usepackage{comment}

\crefname{equation}{}{}

\begin{document}

\title{Quantum Extreme Reservoir Computing for Phase Classification of Polymer Alloy Microstructures}

\author{Arisa Ikeda}
 \affiliation{Graduate School of Science and Technology, Keio University, 3-14-1, Hiyoshi, Kohoku-ku, Yokohama, Kanagawa
 223-8522, Japan}
 
\author{Akitada Sakurai}%
\affiliation{%
Quantum Information Science and Technology Unit, Okinawa Institute of Science and Technology Graduate University, Onna-son, Okinawa 904-0495, Japan}%

\author{Kae Nemoto}%
\affiliation{%
Quantum Information Science and Technology Unit, Okinawa Institute of Science and Technology Graduate
 University, Onna-son, Okinawa 904-0495, Japan}%
 
\author{Mayu Muramatsu}
 \email{muramatsu@mech.keio.ac.jp}
 \affiliation{Department of Mechanical Engineering, Keio University, 3-14-1, Hiyoshi, Kohoku-ku, Yokohama, Kanagawa
 223-8522, Japan}

\date{\today}
\begin{abstract}
Quantum machine learning (QML) is expected to offer new opportunities to process high-dimensional data efficiently by exploiting the exponentially large state space of quantum systems. In this work, we apply quantum extreme reservoir computing (QERC) to the classification of microstructure images of polymer alloys generated using self-consistent field theory (SCFT). While previous QML efforts have primarily focused on benchmark datasets such as MNIST, our work demonstrates the applicability of QERC to engineering data with direct materials relevance. Through numerical experiments, we examine the influence of key computational parameters—including the number of qubits, sampling cost (the number of measurement shots), and reservoir configuration—on classification performance. The resulting phase classifications are depicted as phase diagrams that illustrate the phase transitions in polymer morphology, establishing an understandable connection between quantum model outputs and material behavior. These results illustrate QERC performance on realistic materials datasets and suggest practical guidelines for quantum encoder design and model generalization. This work establishes a foundation for integrating quantum learning techniques into materials informatics.

\end{abstract}

\maketitle
\section{Introduction}

Quantum systems possess exponentially large Hilbert spaces, enabling high-dimensional feature representations even with a modest number of qubits. This property has motivated the development of quantum machine learning (QML) models that exploit quantum dynamics to map input into extremely high-dimensional space~\cite{wilson2019,Noori2020,Schuld2021,Xiong2025}. Among these QML models, quantum reservoir computing (QRC)~\cite{Nakajima2019,Chen2019aa, martinez2021dynamical, Bravo2022, Pfeffer2022Hybrid, Suzuki2022Natural, Xia2022, domingo2022optimal, dudas2023quantum, gotting2023exploring} and quantum extreme learning machines (QELM)~\cite{fujii2017harnessing, govia2021quantum,sakurai2022quantum,De2025Harne,Sakurai2025Boson,Lorenzis2025_2} have attracted growing attention. In these architectures, the quantum component acts as a fixed, untrained reservoir, while only the classical output layer is optimized. Such models circumvent the barren plateau problem~\cite{McClean2018Barren,Qi2023barren}—a major challenge in variational circuit training—and can leverage the complexity quantum dynamics accommodates for learning tasks. Given the simplicity of the quantum part and the low control requirements that enable quantum device dynamics to be directly used as computational resources, these models have been regarded as one of the use cases for Near-Intermediate-Scale Quantum (NISQ) devices.

Early implementations of QRC and QELM were limited to low-dimensional regression problems due to constraints on input encoding~\cite{Chen2019aa,Nakajima2019,Schuld2021, martinez2021dynamical, Pfeffer2022Hybrid,Suzuki2022Natural,Bravo2022}. However, recent advances combining classical preprocessing and dimensionality reduction with quantum reservoirs have extended their applicability to high-dimensional data such as images~\cite{sakurai2022quantum, kornjaca2024large,Senokosov_2024,De2025Harne,Sakurai2025Boson}. Quantum models have now achieved promising performance on standard benchmarks, including MNIST~\cite{lecun1998gradient}, Fashion-MNIST~\cite{xiao2017fashion}, and KMNIST~\cite{clanuwat2018deep}, demonstrating that QML can complement or enhance classical neural approaches in image-based recognition tasks~\cite{hur2022quantum, shen2024classification}.

Despite this progress, applications of QML to scientific and engineering data remains limited. Materials and manufacturing datasets, often rich in structure and directly linked to physical properties, offer a valuable testbed for assessing the potential of quantum learning models in real-world computational science. 

In this work, we apply quantum extreme reservoir computing (QERC) to phase classification of microstructure images of polymer alloys generated by self-consistent field theory (SCFT)~\cite{matsen1994stable, arora2016broadly, drolet1999combinatorial}. Polymer alloys—composed of multiple polymer species designed to tailor mechanical, optical, or thermal properties—exhibit complex microphase-separated structures whose morphology strongly influences macroscopic behavior. Accurate classification of these microstructures is essential for understanding phase transitions and guiding materials design.

We use QERC to classify these SCFT-generated microstructure images and systematically examine computational factors such as the number of qubits, number of shots, and reservoir configuration. The classification results are visualized in phase diagrams reconstructed from quantum model outputs, allowing for direct interpretation of model performance and phase boundary prediction. This visualization provides insight into how quantum models generalize across materials parameter spaces and how encoder design affects the interpretability of quantum feature representations.

\begin{figure*}[thb]
    \centering
    \includegraphics[width=1.9\columnwidth]{ 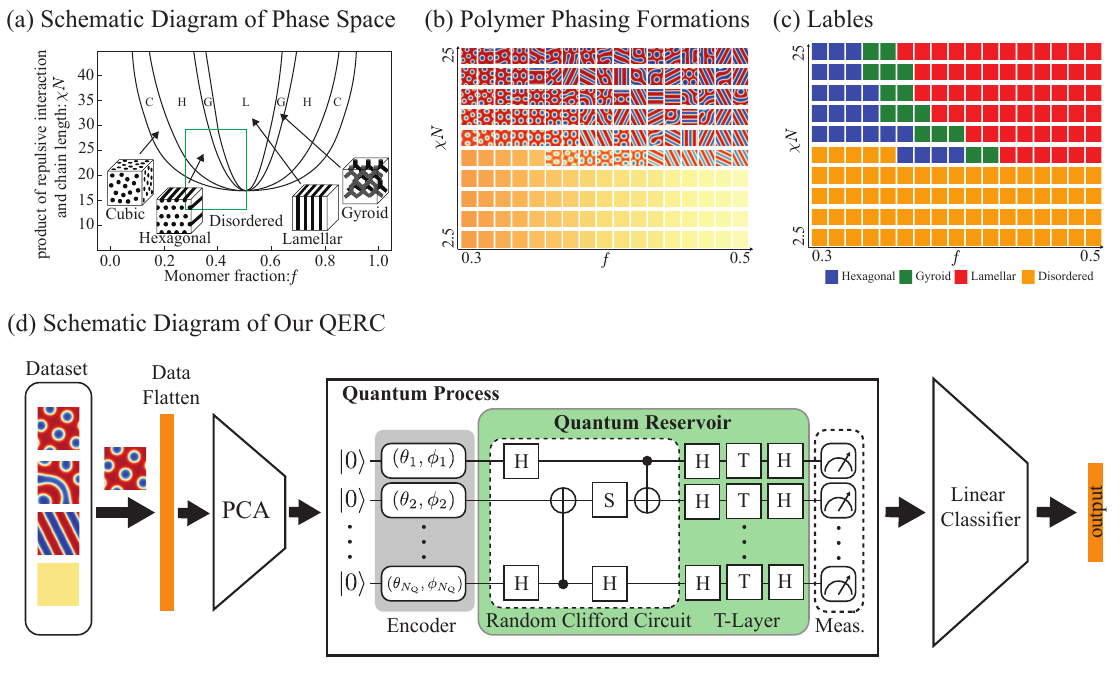}
    \caption{Phase diagram for microstructures of polymer alloys and conceptual diagram of QERC. (a) Schematically represents the differences in microstructure due to the interaction $f$ between segments and the product of the $\chi$ parameter and the degree of polymerization $N$. (b) shows an example of a phase diagram based on the microstructure of a polymer alloy, (c) indicates the correct labels in the phase diagram. (d) is a conceptual diagram of QERC.}
    \label{purpose_phasediagram}
\end{figure*}

Through this work, we aim to (i) evaluate the feasibility and performance of QERC on realistic engineering datasets, (ii) establish a benchmark dataset and methodology for future QML applications in materials informatics, and (iii) provide a foundation for developing interpretable, physics-informed quantum learning models. By demonstrating the utility of QERC in polymer microstructure classification, this work helps bridge the gap between quantum learning theory and practical materials engineering.

\section{Methods}
\label{Methods}
In this section, we describe how the microstructure image dataset of the polymer alloy is prepared and categorized using the QERC model.

\subsection{Classification setup and dataset creation}
Here, we briefly describe the classification problem of polymer alloy microstructures and the method for preparing the dataset. 
Polymer alloys are multicomponent materials composed of two or more types of polymers. 
These different polymers do not necessarily blend completely, and under certain conditions, a heterogeneous state called phase separation can occur. 
Microphase separation refers to phase separation at the microscopic scale of tens of nanometers, forming what is known as a microphase-separated structure. 
Examples of these structures, which consist of two types of polymers, include hexagonal, gyroid, and lamellar structures, as shown in Fig.~\ref{purpose_phasediagram} (a). 
Structures that have not undergone phase separation are called disordered. 
These microstructures can be understood using a phase diagram with two parameters, $(f,\chi N)$, as shown in Fig.~\ref{purpose_phasediagram} (b). 
Here $f$, $\chi$ and $N$ are the volume fraction, the repulsive interaction coefficient and the total number of segments. 
This study treats microstructure as image data and frames a classification task that predicts the corresponding structural type (four structures) in the micro scale. 
From this point onward, if there is no risk of misunderstanding, we will simply refer to this microstructure image as "the image".
We consider a classification of microphases to determine the microphase-separated structure through the image classification using the quantum machine learning model shown in Fig.~\ref{purpose_phasediagram}(d).

The images (cross-sections of microstructures) in this study are generated by the SCFT simulation.
Since microstructures depend on initial state, even when using the same parameters, different initial states will generate microstructures with different appearances. 
This means one microstructure corresponds to one initial random seed. Therefore, considering the reproducibility of numerical calculations, this study generated one microstructure from one random number seed.
To generate different types of phases, the two parameters $(f,\chi)$ are divided within a specific range as follows, and the phase diagram is divided into a grid. 
The range of $f$ is $[0.3, 0.5]$ with a step size of $0.0125$, and the range of $\chi$ is $[0.1, 1.0]$ with a step size of $0.1$.
Where the degree of polymerization $N$ is fixed at $25$.
For the parameters used in this simulation ($f$, $\chi$, $N$, etc.) and detailed content, refer to the Appendix.

Because of considering the supervised model, we need to prepare labels (answers) for the training images. 
Training in our scheme requires training data with labels (answers).  
Our polymer alloys exhibit four microphases hexagonal, gyroid, lamellar, and disordered, and the boundaries of these microphases are not trivial.  
Hence, we first employ the boundaries of the three microphases: hyxagonal, gyroid and lamellar phases determined by the SCFT simulations by Matsen et al.~\cite{matsen1996unifying}.  
Both the linear and spline interpolations based on Table 1 in Ref.~\cite{matsen1996unifying} give the same boundaries.  
The boundaries of disordered and other structures can be determined based on the theory proposed by Fredrickson et al.~\cite{fredrickson1987fluctuation}. See the Appendix for the details.

Finally, we explain the composition of the training and test datasets.
As mentioned above, the parameter space is grid-based and consists of 170 points in total.
For the training data, 24 images were generated at each macro state (grid point in the parameter space) using a different random seed. Similarly, for the test data, 10 images were generated at each point. Therefore, there are 4,080 training images and 1,700 test images.
The number of images included in the training data differs for each of the four structure types, and the dataset does not include any bias mitigation. 

\subsection{Visualization of phase diagrams based on classification results}
\label{phasediagram_evaluation_method}

\begin{figure}[thb]
    \centering
    \includegraphics[width=0.9\columnwidth]{ 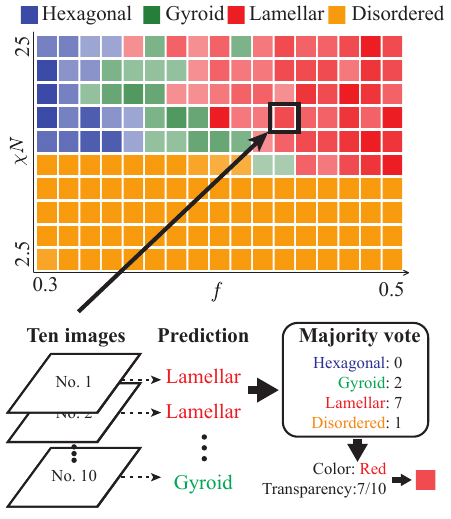}
    \caption{Visualization of the phase diagram based on classification results:
For each grid point in the parameter space, ten test images were generated using different random seeds. The class label for each grid point was determined by majority voting on the QERC classification results, and the corresponding color was assigned accordingly. The transparency was adjusted in proportion to the majority ratio.}
    \label{PhaseVis}
\end{figure}

Since the QERC model is used to predict phase diagram through the classification of  microtstucture images of polymer alloys, the phase diagram prediction as well as the classification accuracy are the primary metric for the performance.
Additionally, within the current problem setting, the phase diagram predicted by the model can also serve as an evaluation indicator. We see this phase diagram visualization is the classification in the parameter space. 
By visualizing this two-dimensional result, it becomes possible to access information, such as the boundaries between classes, that cannot be captured solely by accuracy metrics. In this section, we describe the method used to construct a phase diagram from the classifier's outputs.

To reconstruct the phase diagram from the classification results, we assign colors to each grid point. 
The colors for each grid point is done statistically by using the grid coordinate to generate multiple images for each grid point.
At each grid point, 24 images are generated as the training datase and 10 images for the test dataset, then the color is assinged to the grid point by majority vote, as shown in Fig.~\ref{PhaseVis}.
We also vary the transparency to indicate the majority ratio, so that we visualize multi-modal or uncertain information in the inference.

\subsection{Quantum Extreme Reservoir Computing (QERC)}
This section describes the QERC model, the quantum machine learning model used in this study, and its process.
The basic process is similar to that described in Ref.~\cite{sakurai2022quantum}, where the dimension of the image vector (microstructure image) is much larger than the number of qubits $N_\mathrm{Q}$ in the quantum device being used.

\begin{enumerate}
\item \textbf{Dimensionality reduction and rescaling}:
The input vector $\boldsymbol{x}$ is compressed to a $2N_\mathrm{Q}$-dimensional vector using principal component analysis (PCA) for the training dataset.
Subsequently, this vector is rescaled from 0 to $\pi$ following the same procedure described in Ref.~\cite{sakurai2022quantum}. The rescaled vector is also denoted as $\boldsymbol{x}$.

\item \textbf{Encoding into quantum states}:
The compressed data $\boldsymbol{x}$ is encoded into a quantum state following the procedure in Refs.~\cite{La2020Rbus,sakurai2022quantum,De2025Harne}.
Using single qubit rotations, we encode the data $x_l$  to the quantum state as
\begin{equation}
|\Psi (\boldsymbol{x})\rangle_\mathrm{E} = \prod_{l=1}^{N_\mathrm{Q}}\left(\cos\left(\frac{\theta_l}{2}\right)|0\rangle + e^{i\phi_l}\sin\left( \frac{\theta_l}{2} \right)|1\rangle \right),
\label{eq:single-qubit}
\end{equation}
where  $\theta_l = x_l$ and $\phi_l = x_{N_\mathrm{Q+1}}$ $(l\in (1,2,\cdots,N_\mathrm{Q}))$.

\item \textbf{Quantum reservoir processing}:
The encoded state $|\Psi(\boldsymbol{x})\rangle_\mathrm{E}$ is transformed by the quantum reservoir (characterized by the unitary operator $\hat{U}$) into the final state $|\Psi(\boldsymbol{x})\rangle_\mathrm{F}$ :$|\Psi(\boldsymbol{x})\rangle_\mathrm{F}=\hat{U}|\Psi(\boldsymbol{x})\rangle_\mathrm{E}$

\item \textbf{Linear classifier}:
We obtain the probability for each computational basis to be measured.
When the number of qubits is $N_\mathrm{Q}$, the computational basis has $2^{N_\mathrm{Q}}$ components.
These are normalized as vectors and input into a subsequent neural network (NN)~\cite{haykin2009}.
\end{enumerate}

The unitary operator $\hat{U}$ in the step 3 in the above procedure still has to be determined.  In this study we employ "Clifford+T" to construct $\hat{U}$~\cite{sakurai2025simple}.   A "Clifford+T" quantum circuit is a random sequence of Clifford operators followed by a T-gate layer, which is easily implemented on gate-base quantum processors~\cite{Mooney2021whole,Pphilipp2022Reali} such ones provided by IBM~\cite{javadi2024quantum}.   Clifford operators are uniformly sampled from the Clifford operator set. Since the T-gate is a phase gate, it is ineffective for measurement on its own, so it is placed between H-gates. 
We performed quantum circuit simulations using Qiskit~\cite{javadi2024quantum}.
The classical NN was optimized using the cross-entropy loss function and the AdaGrad algorithm~\cite{Duchi2011Adap}, as described in Ref. ~\cite{sakurai2025simple}.

\section{Results}
\label{Results}
We present the primary results  of the image analysis through the QERC classification in this section.

\subsection{Classification accuracy vs the number of qubits}
\begin{figure}[tb]
    \centering
    \includegraphics[width=0.9\columnwidth]{ 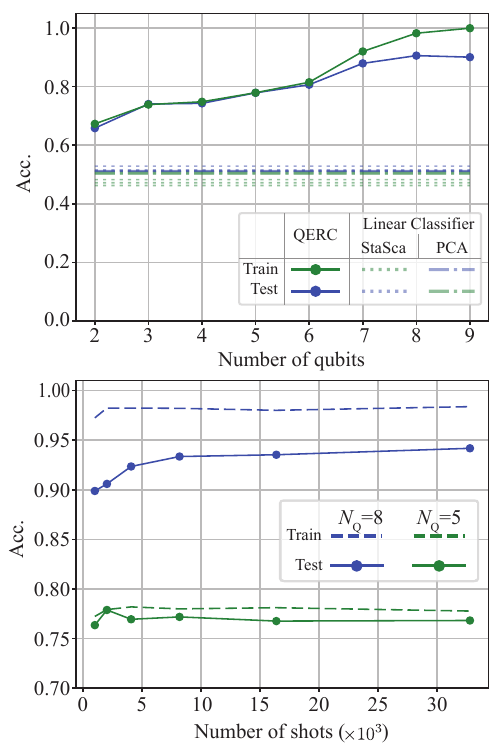}
    \caption{Difference in accuracy based on qubit count. Solid lines show results using QERC, dashed lines show results using a single-layer neural network (NN) without a quantum reservoir. Since random number effects significantly impact linear classification, results vary easily even without changing conditions; thus, results were verified three times each.}
    \label{fig:numberofqubitsshots_largedata}
\end{figure}

To evaluate the capability of the QERC model and its  resource requirements, we analyse the classification performance for the different quantum reservoir sizes from 2 to 9 qubits.  The results are shown in Fig.~\ref{fig:numberofqubitsshots_largedata}.
The number of shots is set to 2,048 to ensure the probability distribution could be reconstructed with sufficient precision.

The horizontal axis represents the number of qubits, and the vertical axis represents the classification accuracy achieved by QERC.
The solid green line represents training accuracy, and the solid blue line represents the test accuracy.
The dashed line represents the classification results obtained using a single-layer NN without a quantum reservoir.
Linear separation is performed on both the dataset processed with Scikit-learn's StandardScaler~\cite{scikit-learn} and the dataset processed with PCA.
The results show that the accuracy improves significantly in comparison to the results obtained using only NN, due to the effect of the quantum reservoir.
The accuracy improves gradually up to 6 qubits and then rapidly thereafter.
The test accuracy starts to convert around 7 qubits.
Therefore, it is considered that this dataset can produce sufficient classification results with approximately 7 qubits.

The output (bit string measured) of a quantum circuit is probabilistic, and the reconstructed probability distribution exhibits statistical variation. 
Consequently, the results of QERC may change depending on the number of shots.
Fig.~\ref{fig:numberofqubitsshots_largedata} shows the results obtained by varying the number of shots to evaluate this effect.
For 5 qubits or fewer, the states are primarily classified into two types: lamellar and disordered. 
The hexagonal and gyroid states are predicted to be mostly lamellar.
Some predictions for the hexagonal phase appear at 6 qubits, but regions predicted to be lamellar and disordered dominate.
The region predicted as hexagonal is also seen to be incorrect when compared to Fig.~\ref{purpose_phasediagram} (b). 
The prediction of the gyroid appears at 7 qubits, and the phase diagram is reproduced correctly at 8 qubits.

\begin{figure}[tb]
    \centering
    \includegraphics[width=0.48\textwidth]{ 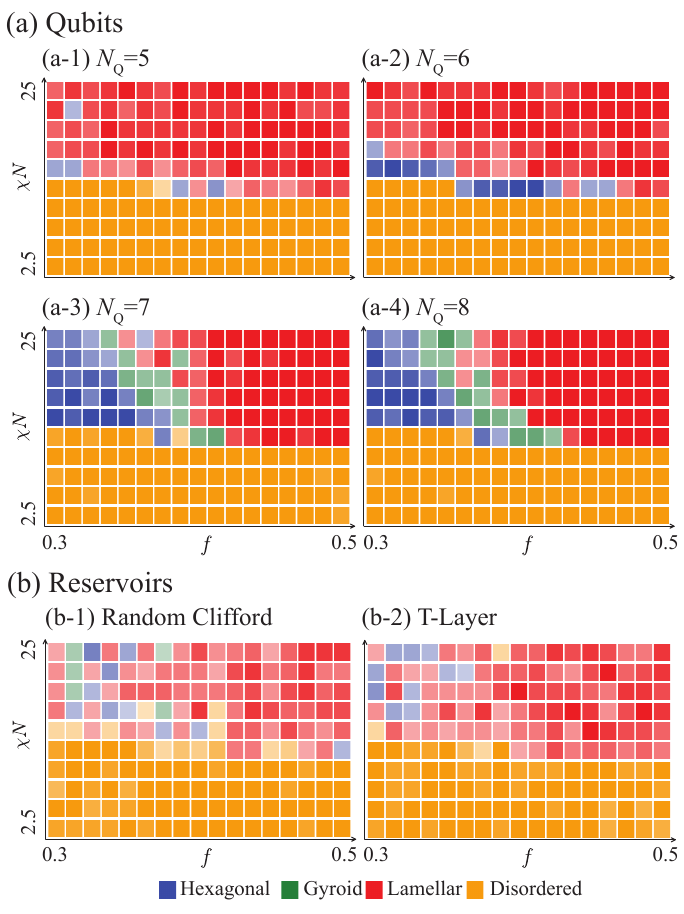} 
    \caption{Visualization of phase diagrams based on QERC predictions: (a) The number of qubits is 5 to 8. The number of shots is 2,048. The reservoir is random Clifford + T-layer. (b) The reservoir is random Clifford or T-layer. The number of qubits is 8 and the number of shots is 2,048.}\label{fig:qubit_majority_comparison}
\end{figure}

Next, to investigate the classification tendency of QERC for different numbers of qubits, we constructed predicted phase diagrams from the test dataset following the procedure described in Methods (1)~\ref{phasediagram_evaluation_method}.
The predicted phase diagrams obtained by QERC for 5 to 8 qubits are shown in Fig.~\ref{fig:qubit_majority_comparison}.
As shown in Fig.~\ref{fig:qubit_majority_comparison}(a), for five qubits or fewer, the results are mainly classified into two categories: lamellar and disordered, while the hexagonal and gyroid phases are almost entirely predicted as lamellar.
At six qubits, a small number of hexagonal predictions begin to appear, but the regions predicted as lamellar and disordered remain dominant.
In addition, a comparison with Fig.~\ref{purpose_phasediagram}(b) reveals that the regions predicted as Hexagonal are actually misclassified.
When the number of qubits increases to seven, predictions corresponding to the gyroid phase start to emerge, and at eight qubits, the predicted phase diagram is largely consistent with the ground truth.

\subsection{Verification of the reservoir parts using only random Clifford circuits or only T-gates}
In the numerical calculations presented so far, we employed the “Clifford+T” quantum circuit—composed of random Clifford gates and a single layer of T gates—as the quantum reservoir.
Previous studies have reported that removing the T-gate layer from the circuit significantly degrade the performance.
Therefore, in this study, we also examined two independent quantum reservoirs: one consisting solely of a random Clifford circuit and the other of a T-gate layer. The results are shown in Fig.~\ref{fig:qubit_majority_comparison}(b).
As in the previous analysis, the number of shots is set to 2,048.
When using only the random Clifford circuit or only the T-gate layer, the Disordered phase is mostly classified correctly, whereas other structural phases could not be properly identified.
These results indicate that, for our dataset, both the random Clifford circuit and the T-gate layer are essential for achieving adequate QERC performance.

\section{Discussions}
\subsection{Verification of factors contributing to accuracy dependence on the number of qubits}
As seen in the results , the classification performance of QERC improves dramatically as the number of qubits increases from 6 to 7.
Considering the characteristics of QERC, improvements in classification performance with increasing the number of qubits can be attributed to two factors: an increase in the number of generated feature mappings (increased representational capacity of the model) and an increase in the number of PCA components used in the encoder (increased input information).

\begin{figure}[tb]
    \centering
    \includegraphics[width=0.45\textwidth]{ 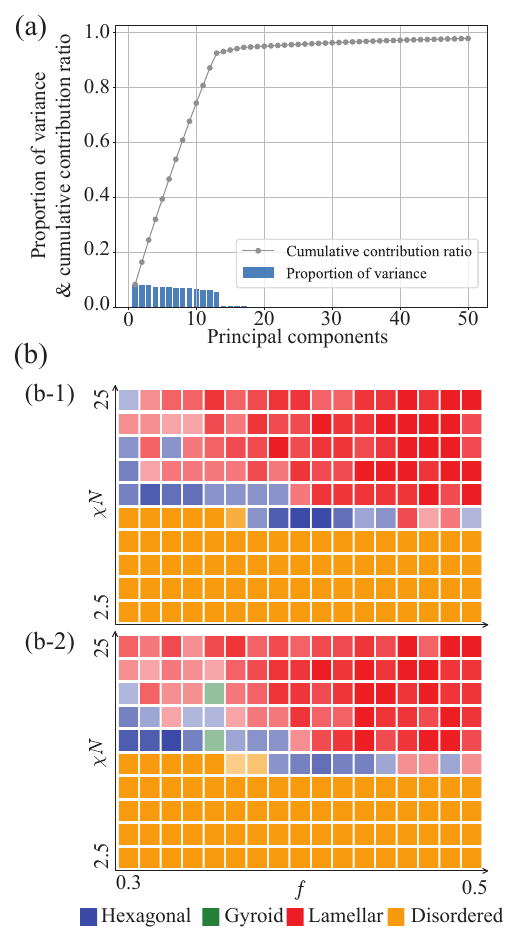} 
    \caption{Visualization of contribution rates and cumulative contribution rates after PCA application and phase diagrams: (a) PCA contribution rate and cumulative contribution rate. (b) Phase diagram : (b-1) 7 qubits result, encoding components 1 to 12 and components 15 to 16; (b-2) 8 qubits result, encoding components 1 to 12 and components 17 to 20.}\label{fig:PCA_contribution}
\end{figure}

To answer which of these two factors is the primary, we first check the contribution rate of PCA for this dataset.
Fig.~\ref{fig:PCA_contribution} shows the contribution rates and cumulative contribution rates of the principal components after dimension reduction with PCA. 
The first 13 principal components have large contribution rates, while those beyond that point exhibit a sharp decrease in contribution rates. 
The contribution rates of components 1 through 13 are roughly equal, so these components can be considered equally important components (information) that represent the image. 

The encoder we used encodes the first to $N_\mathrm{Q}$th components of the $N_\mathrm{Q}$ qubits to angle $\theta_l$, and the $(N_\mathrm{Q}+1)$th to $2N_\mathrm{Q}$th components to phase $\phi_l$.
Therefore, only up to 12 components are encoded in 6 qubits, and the 13th component is not included.
The 7th qubit is necessary to encode the 13th component.
To investigate the impact of excluding the 13th component during encoding, we shifted components in both the 7-qubit and 8-qubit encodings to intentionally exclude the 13th component.
The results are shown in Fig.~\ref{fig:PCA_contribution}.
The classification performance of the 7 and 8 qubit results is significantly lower than that of the standard encoding results.
In particular, it is found that the lack of the 13th component affects the classification of hexagonal and gyroid.
From these results, it is shown that in encoders using PCA, the lack of components with relatively high contribution rates significantly impacts the model's classification performance.

The above analysis confirms that the classification performance varies significantly depending on how the principal components of the encoded images are selected. 
In the present experiment, the contribution ratios indicated that the first thirteen components were particularly important; however, such a clear tendency cannot necessarily be expected for general datasets. 
In learning models that use dimensionality reduction, including QERC, it is challenging to determine in advance which and how many principal components or features from techniques like auto-encoders~\cite{Lorenzis2025_2, De2025Harne} are needed, and the compression dimension should be set with some margin. 
While the number of components that can be input into a quantum system is limited to $2\times N_\mathrm{Q}$, and when $N_\mathrm{Q}$ is fixed, it is not obvious if it is possible to increase the number of input elements. 
Therefore, in dimensionality-reduction-based learning using quantum models, balancing the selection of principal components with the constraint imposed by the number of qubits becomes a critical factor determining overall performance. 
Furthermore, when revisiting the results obtained with different quantum reservoirs, it was observed that using a Random Clifford circuit or a T-gate layer as the reservoir did not yield sufficient classification performance, even though the eight-qubit system contained enough encoded components (see Fig.~\ref{fig:qubit_majority_comparison} (b)). 
This finding suggests that, in QERC, not only sufficient input information but also effective information scrambling by an appropriate quantum reservoir is essential. The present results indicate that both the Random Clifford circuit (entangling gates) and the T-gate layer (phase gates) play complementary roles in realizing meaningful information mixing within the quantum reservoir, thereby contributing to improved learning performance.

\subsection{Evaluation of generalization performance of QML models on parameter space}
\begin{figure}[tb]
    \centering
    \includegraphics[width=0.40\textwidth]{ 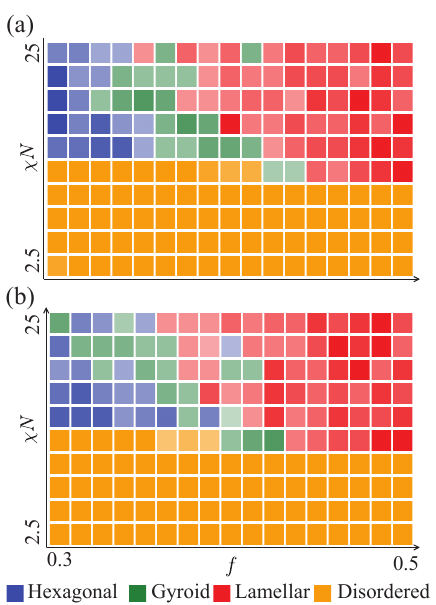} 
    \caption{
     Evaluating generalization performance in parameter space: (a) Classification results after bias mitigation with downsampled training data from QERC with 8 qubits and 2,048 shots ($\chi N = 2.5, 7.5, 12.5, 17.5, 22.5$ omitted). (b) Classification results after bias mitigation with downsampled training data from QERC with 8 qubits and 2,048 shots ($\chi N = 5.0, 10, 15, 20, 25$ omitted).)
    } \label{fig:8qubit_biasmodelcheck}
\end{figure}

Thus far, images for the training and testing data have been generated using the same grid points in the parameter space, but different random number seeds have been used.
This suggests that the results so far have evaluated the generality of QERC in relation to the initial randomness (random seed).

The dataset for this study was generated from numerical simulations based on the theorical model. 
It allows for a certain degree of freedom, such as parameter selection, enabling the creation of different datasets depending on the performance evaluation of the learning model.
Therefore, we generated the following dataset and evaluated the generalization performance of QERC's parameters.

In the new dataset, the test data uses grid points in the parameter space as before, but the training data uses a grid that has been downsampled.
Specifically, the training data was generated by two methods of data thinning to prepare the dataset.
The selection of parameter $f$ is common to both datasets, excluding data points on the $f=0.3125+0.025n$, where $n \in \{0,1,\dots,7\}$.
By contrast, the selection of the parameter $\chi N$ differs: the first dataset is based on $\chi N=2.5+5.0n\,(n\in\{0,1,\dots,4\})$, while the second is based on $\chi N=5+5.0n\,(n\in\{0,1,\dots,4\})$, both representing data sets with thinned points.

Furthermore, to prevent imbalance in the number of images per class during training dataset generation, we performed bias mitigation by sampling an equal number of images from each of the four class sets.
The bias mitigation means the process of equalizing the number of samples in each class.
As a result, the training dataset contains 344 images, while the test dataset remains at 1,700 images as before.
In contrast to previous datasets, the test dataset is larger, while the training dataset is smaller.

The results of the QERC optimized on two training dataset for the test dataset are shown in Fig.~\ref{fig:8qubit_biasmodelcheck}.
Compared to the results from the original training dataset, the results from the new training data show an increase in classification errors.
However, considering that the training dataset is one-fifth the size of the test dataset, the test results can be considered generally satisfactory.
These results indicate that QERC possesses sufficient generalization performance even with respect to the parameter space.
Thus, the dataset based on physical numerical calculations used in this study can be flexibly configured according to the evaluation of the learning model.

\section{Conclusion}
In this work  we performed classification of polymer-alloy microstructure images generated by self-consistent field theory (SCFT) using the quantum extreme reservoir computation (QERC), a type of quantum machine learning (QML) model. Our results demonstrate that high-precision classification can be achieved through mapping into higher-dimensional feature spaces via quantum reservoirs, requiring only about seven qubits of physical resources. We further examined the influence of various computational parameters, such as the number of measurement shots, on classification performance. The QERC model was shown to perform effectively even in the boundary regions between different classes.

By visualizing the phase diagram, we confirmed that each phase structure is accurately reproduced within the parameter space. This visualization not only reflects the classification accuracy but also provides detailed insights into the internal behavior of QML models—offering valuable guidance for future encoder and model design as well as for improving interpretability. Moreover, the physically grounded, simulation-based dataset employed here serves as a useful benchmark for systematically assessing the generalization and robustness of learning models, owing to its flexible control over data-generation conditions.

Overall, these findings highlight the potential of QML approaches such as QERC for the analysis of engineering datasets and underscore their promise as practical tools for future industrial applications.

\begin{acknowledgments}
We are grateful to Yutaka Oya for valuable and insightful discussions, and to W. J. Munro for helpful comments on the manuscript.
This work was supported by the Japan Science and Technology Agency (JST) through the COI-NEXT Program (Grant No. JPMJPF2221), the FOREST Program (Grant No. JPMJFR212K), and  JSPS KAKENHI under Grant No. 25K21306. 

\textbf{Data availability:}
Data underlying the results presented in this paper are available from the authors upon reasonable request.
\end{acknowledgments}

\appendix
\section{Preparing polymer alloy microstructure images using SCFT}
We provide a detailed description of the microstructure generation process here to enable reproduction of the dataset used in this study.

Polymer alloys form microstructures with various ordered structures, as shown in Fig.~\ref{purpose_phasediagram}.
Even when phase separation occurs, if the separated phases cannot grow to large sizes and instead form stable structures remaining at around tens of $\mathrm{nm}$, these are called microphase-separated structures.
Since most microphase-separated structures possess order, this state is called the ordered state, while a state uniformly mixed at the monomer level is called the disordered state. 
This morphology depends on the chain length, the volume fraction of each block, and the inter-segment interaction parameter $\chi$.
Where, the $\chi$ parameter corresponds to the miscibility between different chemical structures within each block chain, and a larger value of $\chi$ indicates lower miscibility.
Fig.~\ref{purpose_phasediagram} shows the volume fraction of polymer A in the A{--}B diblock copolymer on the horizontal axis and the product of the $\chi$ parameter representing inter-segment interactions and the degree of polymerization $N$ on the vertical axis.
These microstructures influence the macroscopic mechanical properties of materials\cite{hiraide2023inverse, hiraide2024development}.
Microstructures change continuously with parameters, but the phase diagram determines the ordered structure.
Among phase-separated microstructures, the gyroid is a three-dimensional co-connected network structure based on a surface that minimizes the interfacial area.
It was discovered in 1986 by Thomas et al.~\cite{thomas1986ordered} and in 1987 by Hasegawa et al.~\cite{hasegawa1987bicontinuous} as a new morphology exhibiting a co-continuous structure called OBDD (ordered biscontinuous double diamond lattice) with an intermediate composition between cylindrical and layered structures. 

As a dataset for QERC, we prepared two-dimensional microstructures of polymer alloys using SCFT simulations~\cite{matsen1994stable}. 
Computational simulations of the microphase-separated structure model for block copolymers have become possible, and Bates et al.~\cite{matsen1996unifying, bates1999block} theoretically demonstrated the prediction of phase diagrams. 
Using density functional theory with a self-consistent field approach, the phase diagram of an A{--}B diblock copolymer melt can be depicted.
In SCFT simulations, a single polymer chain is coarse-grained using a spring-bead model where segments are connected by bonds.
The interaction between numerous polymer chains was approximated as a single polymer chain in a mean-field environment.
We analyzed the system by changing the seed value of the random number generator that determines the initial state, and generated multiple phase diagrams as a dataset.
In SCFT simulations, parameters are repeatedly updated until the following three equations become consistent under non-compression conditions.
The mean field, path integral, and segment concentration are expressed as follows : 
\begin{subequations}
\begin{equation}
    V_i(\boldsymbol{r})= \sum_{j=A,B} \chi_{ij}\phi_j(\boldsymbol{r})+\gamma(\boldsymbol{r}),
\end{equation}
\begin{equation}
    \frac{\partial Q_i}{\partial s} = \frac{b^2}{6}\nabla^2 (Q_i\boldsymbol{r}_0, s_0;\boldsymbol{r},s)-\beta V_i(\boldsymbol{r})Q_i(\boldsymbol{r}_0,s_0,;\boldsymbol{r}, s),
\end{equation}
\begin{equation}
    \phi_i(\boldsymbol{r})= M\frac{\int_{0}^{N_i}ds\int d\boldsymbol{r}_0\int d\boldsymbol{r}_{N_i}Q_i(\boldsymbol{r}_0,s_0;\boldsymbol{r},s)Q_i(\boldsymbol{r},s;\boldsymbol{r}_{N_i},s)}{\int d\boldsymbol{r}_0 \int d\boldsymbol{r}_{N_i}Q_i(\boldsymbol{r},s;\boldsymbol{r}_{N_i},N_i)},
\end{equation}
\end{subequations}
where $V_i(\boldsymbol{r})$is the mean field, $\boldsymbol{r}$ is the position, $\chi_{ij}$ is the interaction coefficient between polymer $i$ and polymer species $j$, $Q_i$ is the path integral, $s$ is the segment number, $\phi_i$ is the segment concentration, $\gamma$ is the constraint force due to the incompressibility condition, $b$ is the bond length, $\beta=1/{k_BT}$, $k_B$ is the Boltzmann constant, $T$ is the absolute temperature, $M$ is the total number of chains of polymer species $K$ in the system, and $N_i$ is the total number of segments.
Table~\ref{SCF_condition} shows the conditions used for the SCFT simulations. 
The range and interval of volume fraction $f$ and repulsive interaction $\chi$ were determined to enable the observation of four distinct structures: hexagonal, gyroid, lamellar, and disordered.

\begin{table}[htb]
    \caption{
        \textbf{The conditions for SCFT simulations}{\textmd{}}
    }
    \label{SCF_condition}
   \footnotesize
    \begin{tabular}{l l}\hline
        Number of iterations & 300,001\\
        Number of joints $s_n$ & 500\\
        Bond length $d_S /\text {nm}$ & 0.05\\
        Total number of segments $N$ & 25\\
        System size & $16\times16$\\
        Number of lattice points & $64\times64$\\
        Boundary condition & periodic boundary condition\\
        Volume fraction $f$ & $\left[ 0.3, 0.5 \right] $ (increments of 0.0125)\\
        Repulsive interaction coefficient $\chi$ & $\left[ 0.1, 1.0 \right] $ (increments of 0.1)\\ \hline
    \end{tabular}
\end{table}

\section{Labeling of microstructures}
We describe in detail the method for labeling microstructures.
The labeled phase diagram is shown in Fig.~\ref{purpose_phasediagram} (c).
The labels are shown in Fig.~\ref{purpose_phasediagram} (c), corresponding to the same locations in phase diagram Fig.~\ref{purpose_phasediagram} (b).
The boundaries of the hexagonal, gyroid, and lamellar phases were determined using the results of SCFT simulations by Matsen et al.\cite{matsen1996unifying}.
The boundaries were determined by linear interpolation and spline interpolation using Table 1 from the Ref.~\cite{matsen1996unifying}. 
The results of linear interpolation and spline interpolation were the same.
The boundary between disordered and other structures was determined by ref. to the theory proposed by Fredrickson et al.~\cite{fredrickson1987fluctuation}. 
From the ref. \cite{fredrickson1987fluctuation}, spinodal is expressed as follows :
\begin{equation}
    (\chi N)_{\rm{spinodal}} = \frac{F(x^*,f)}{2},
\end{equation}
where, from the Flory parameter calculation formula, $F$ is expressed as follows : 
\begin{subequations}
\begin{equation}
\begin{aligned}
F(x,f)
= &\; g(1,x)\Big/ \Bigl(
      g(f,x)g(1-f,x) \\[2pt]
&\qquad
      - \tfrac{1}{4}\bigl[\,g(1,x)-g(f,x)-g(1-f,x)\,\bigr]^2
    \Bigr),
\end{aligned}
\end{equation}
\begin{equation}
g(f,x)=2\,(fx+e^{-fx}-1)/x^2,
\end{equation}
\end{subequations}
where $f$ is the volume fraction of polymer A in the A--B diblock copolymer and $x$ represents the position.
The boundaries were determined by linear interpolation and spline interpolation using Table 1 from the literature~\cite{fredrickson1987fluctuation}. 
The results for linear interpolation and spline interpolation were also the same. 
This procedure labels them into four classes: hexagonal, gyroid, lamellar, and disordered.

\end{document}